\newcommand{\beq}{\begin{equation}}
\newcommand{\eeq}{\end{equation}}
\newcommand{\bea}{\begin{eqnarray}}
\newcommand{\eea}{\end{eqnarray}}

\newcommand{\gsim}{\lower.7ex\hbox{$\;\stackrel{\textstyle>}{\sim}\;$}}
\newcommand{\lsim}{\lower.7ex\hbox{$\;\stackrel{\textstyle<}{\sim}\;$}}

\newcommand{\mrm}{\mathrm}

\documentclass[aps,prev,twocolumn,preprintnumbers,floatfix,nofootinbib]{revtex4-1}
\usepackage{graphicx}
\usepackage{epstopdf}
\usepackage{mathrsfs}
\usepackage{amssymb}
\usepackage{verbatim}

\def\stacksymbols #1#2#3#4{\def\theguybelow{#2}
    \def\vp{\lower#3pt}
    \def\sp{\baselineskip0pt\lineskip#4pt}
    \mathrel{\mathpalette\intermediary#1}}

\def\intermediary#1#2{\vp\vbox{\sp
     \everycr={}\tabskip0pt
     \halign{$\mathsurround0pt#1\hfil##\hfil$\crcr#2\crcr
              \theguybelow\crcr}}}

\def\be{\begin{equation}}
\def\ee{\end{equation}}
\def\bea{\begin{eqnarray}}
\def\eea{\end{eqnarray}}

\def\sp{\;\;\;,\;\;\;}

\def\mrm{\mathrm}

\def\lsim{\raise0.3ex\hbox{$\;<$\kern-0.75em\raise-1.1ex\hbox{$\sim\;$}}}
\def\gsim{\raise0.3ex\hbox{$\;>$\kern-0.75em\raise-1.1ex\hbox{$\sim\;$}}}

\def\inbar{\,\vrule height1.5ex width.4pt depth0pt}

\def\IC{\relax\hbox{$\inbar\kern-.3em{\rm C}$}}
\def\IQ{\relax\hbox{$\inbar\kern-.3em{\rm Q}$}}
\def\IR{\relax{\rm I\kern-.18em R}}
 \font\cmss=cmss10 \font\cmsss=cmss10 at 7pt
\def\IZ{\relax\ifmmode\mathchoice
 {\hbox{\cmss Z\kern-.4em Z}}{\hbox{\cmss Z\kern-.4em Z}}
 {\lower.9pt\hbox{\cmsss Z\kern-.4em Z}}
 {\lower1.2pt\hbox{\cmsss Z\kern-.4em Z}}\else{\cmss Z\kern-.4em Z}\fi}

\def\comment#1{}

\def\u1x{U(1)_X}
\newcommand{\nc}{\newcommand}
\nc{\LL}{L}
\nc{\vv}{\tilde{v}}
\nc{\ccdot}{\!\cdot\!}
\nc{\gsm}{G_{SM}}
\nc{\vfive}{\mathbf{5}\oplus\mathbf{\overline{5}}}
\nc{\vten}{\mathbf{10}\oplus\mathbf{\overline{10}}}
\nc{\zhol}{Z^{\rm hol}}
\nc{\xfb}{\,{\rm fb}}

\begin{document}

%
%

\preprint{CERN-PH-TH/2012-109}
\preprint{CPHT-RR 018.0512}
\preprint{FTUAM-12-90}
\preprint{IFT-UAM/CSIC-12-37}
\preprint{LPT--Orsay 12/39}

\vspace*{1mm}

\title{Extra U(1) as natural source of a monochromatic gamma ray line}

\author{Emilian Dudas$^{a,b}$}
\email{Emilian.Dudas@cpht.polytechnique.fr}
\author{Yann Mambrini$^{c}$}
\email{Yann.Mambrini@th.u-psud.fr}
\author{Stefan Pokorski$^{d,e}$}
\email{Stefan.Pokorski@fuw.edu.pl}
\author{Alberto Romagnoni$^{f}$}
\email{Alberto.Romagnoni@uam.es}

\vspace{0.1cm}
\affiliation{
${}^a$ Department of Physics CERN, Theory Division, CH-1211, Geneva 23, Switzerland
}
\affiliation{
${}^b$ CPhT, Ecole Polytechnique 91128 Palaiseau Cedex, France
 }
\affiliation{
${}^c$ Laboratoire de Physique Th\'eorique
Universit\'e Paris-Sud, F-91405 Orsay, France
 }
\affiliation{
${}^d$ Institute of Theoretical Physics, Warsaw University, Hoza 69, 00-681 Warsaw,
Poland}
\affiliation{
${}^e$TUM-IAS, Technische Universitat  Munchen, Lichtenbergstr. 2A,
D-85748 Garching, Germany
 }
 \affiliation{
${}^f$ Departamento $\&$ Instituto de F\'isica Te\'orica UAM/CSIC,
Universidad Aut\'onoma de Madrid,
Cantoblanco,
28049 Madrid, Spain.
}

\begin{abstract}

Extensions of the Standard Model with an extra $U'(1)$ abelian group generically generate
 terms coming from loops of heavy fermions, leading to three gauge boson couplings, in particular $Z'Z\gamma$.
We show that WMAP data constrains the gauge coupling of the group $g_D$ to values comparable with the electro-weak ones,
rather independently of the mass of $Z'$. Moreover, the model predicts a monochromatic $\gamma$-ray line which can
 fit a 130 GeV signal at the FERMI telescope for natural values of the Chern-Simons terms and a dark matter mass around 144.5 GeV.

\end{abstract}

\maketitle

\section*{Introduction}

One of the most important issues in particle physics phenomenology is the nature
and properties of the dark matter in our universe.
 The observations made by the WMAP collaboration \cite{WMAP} show that the matter content of the universe is dark,
making up about 85 \% of the total amount of matter.
On the other hand, the XENON collaboration recently released its constraints
on direct detection of Dark Matter \cite{Aprile:2011ts} excluding large regions of several extensions of the
Standard Model. These constraints makes it plausible that dark matter sits
in a different sector, communicating with our sector through new, weak enough interactions.

\noindent
Neutral gauge sectors with an additional dark $U'(1)$ symmetry in addition
to the Standard Model (SM) hypercharge $U(1)_Y$ and an associated $Z'$ gauge boson
are among the most natural extensions of the SM, and give the possibility
that a dark matter candidate lies within this new gauge sector of the theory.
Extra gauge symmetries are predicted in most Grand Unified Theories (GUTs)
and appear systematically in string constructions. Larger groups than $SU(5)$
or $SO(10)$ allow the SM gauge group and $U(1)'$ to be embedded into bigger GUT groups.
String theory and brane--world $U'(1)$s are special compared to GUT $U'(1)$'s; some of them are hidden, such
that SM particles are uncharged under them.
For a review of the phenomenology of the extra $U'(1)$s generated in such scenarios see e.g.
 \cite{Langacker:2008yv}.
In such a framework, the extra $Z'$ gauge boson would act as a portal between the $dark$ world
(particles not charged under the SM gauge group) and the $visible$ sector.

\noindent
Several papers considered that the key of the portal could be
the gauge invariant kinetic mixing $(\delta/2) F_Y^{\mu \nu} F'_{\mu \nu}$
\cite{Holdom}.
One of the first models of dark matter from the hidden sector with a massive additional
$U'(1)$, mixing with the SM hypercharge through both mass and kinetic mixings
can be found in \cite{Feldman:2006wd}.
The Dark Matter (DM) candidate $\psi_0$ could be the lightest (and thus stable) particle of
this secluded sector. Such a mixing has been justified in
recent string constructions \cite{Cicoli:2011yh,Kumar:2007zza,
Javier,Cassel:2009pu}, but has also been studied within a model independent
approach \cite{Feldman:2007wj,Pospelov:2008zw,Mambrini:2010yp}
or in a supersymmetric extension \cite{SUSYmixing}.

\noindent
However, there exists another possibility for the $Z'$ portal: the diagrams generated
by the Chern-Simons terms, usually related to the mechanism of gauge anomaly cancelation. It has been shown
\cite{LineDudas,Mambrini:2009ad} that these vertices could generate a specific smoking-gun signal for dark matter
searches:  a monochromatic gamma ray line from the Galactic Center \cite{Vertongen:2011mu}.
Moreover, the recent hint for such a line\footnote{At 3.3 $\sigma$ if one takes
into account the look elsewhere effect, but currently unconfirmed by
FERMI collaboration.}
 \cite{Weniger:2012tx,Raidal}
raises the hope and interests for such theoretical extensions of the Standard Model (see also \cite{Chalons:2012hf,Profumo:2012tr,Ibarra:2012dw} for very recent discussions about this subject). The purpose of the present note is to prove that such models naturally accomodate both the WMAP data on dark matter
and the generation of a monochromatic gamma ray line from the Galactic Center\footnote{Very few models can achieve such a signal as its production is
one--loop suppressed. However, this monochromatic ray can be enhanced in other scenarios like SUSY ones \cite{SUSYline}, extra-dimension
constructions \cite{Extraline}, singlet DM \cite{Singletline}, decaying DM \cite{Decayingline}, including a neutrino sector \cite{Decayingseto},
effective DM models \cite{Effectiveline} or inert Higgs doublet DM \cite{Inertline}. Internal Bremsstrahlung \cite{Bremline} can also exhibit a
spectrum similar to the one produced by the emission of a monochromatic $\gamma$-ray line.
 }.

\noindent
This note is organized as follows. After defining the model,
we present the phenomenological consequences and study the parameter space which could
respect WMAP and simultaneously explain a monochromatic gamma-ray line signal from the Galactic Center.

\section*{The model}

\begin{figure}[!h]
    \begin{center}
   \includegraphics[width=3.in]{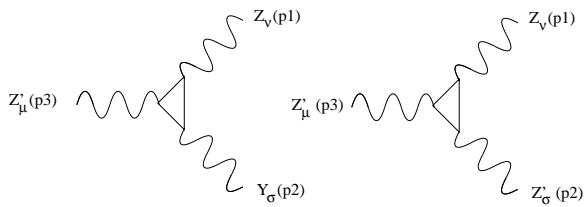}
          \caption{{\footnotesize
Triangle diagrams whose variation generate counter terms of the form Eq.\ref{Eq:chernsimon}
}}
\label{Fig:feynman}
\end{center}
\end{figure}

Gauge invariance is a fundamental condition to ensure renormalizability and quantum consistency
of any extension of the Standard Model. Triangle gauge anomalies cancelation can occur by a consistent field theory content
or by cancelation of  triangle loops by axionic couplings and Chern-Simons terms, via the string theory Green-Schwarz mechanism. At low-energy, remnants of the anomaly cancelation can lead to generalized Chern-Simon terms \cite{Abdk} containing new three gauge boson
couplings. If one extend the Standard Model by an abelian gauge group $U'(1)$, at low-energy the particular Chern-Simons terms
we are interested in is

\beq
{\cal L}_{CS} = \alpha_1 ~ \epsilon^{\mu \nu \rho \sigma} Z'_\mu Z_\nu F^{Y}_{\rho \sigma}
+ \alpha_2 ~ \epsilon^{\mu \nu \rho \sigma} Z'_\mu Z_\nu F'_{\rho \sigma} \ ,
\label{Eq:chernsimon}
\eeq
\noindent
where $CS$ stands for Chern-Simons, $F'_{\rho \sigma} = \partial_{\rho} Z'_{\sigma} - \partial _{\sigma} Z'_{\rho}$
and ($\alpha_1$,$\alpha_2$) are the coefficients
(computable exactly once given the fermionic content of the model) generated by the triangle diagrams
depicted in Fig.\ref{Fig:feynman}. In the loops are running  heavy fermions charged under both $U(1)_Y$ and
$U'(1)$  (when heavy fermion masses are SM gauge invariant  such diagrams are still generated  by higher dimensional operators
  \cite{LineDudas,Antoniadis:2009ze}). Notice that the CS terms (\ref{Eq:chernsimon}) are invariant under electromagnetism\footnote{The broken SM symmetries and the $U(1)'$ are realized in a Stueckelberg phase, as explained in
\cite{LineDudas}. The CS terms can be written in the manifestly gauge invariant way
$ \frac{i}{M^2} \epsilon^{\mu \nu \rho \sigma} D_{\mu} \theta (D_{\nu} H^{\dagger} H- H^{\dagger} D_{\nu} H)  (c_1 F^{Y}_{\rho \sigma}
+ c_2  F'_{\rho \sigma})$, where $D_{\mu} \theta = \partial_{\mu} \theta - g_X Z'_{\mu}$, where $\theta$ is the Stueckelberg
axion absorbed by the $Z'$ gauge boson. Moreover $M$ is a mass scale related to the mass of the heavy fermions. After electroweak symmetry breaking, we recover (\ref{Eq:chernsimon}) with
$\alpha_i \sim c_i v^2/M^2$. We notice here that CS terms are also generic in string constructions, where their gauge non-invariance is
compensated by axionic couplings and triangle loops of light fermions charged under the extra $U(1)'$. If light fermions are present,
the computations of three gauge boson vertices are changed qualitatively, see e.g. \cite{Abdk}, \cite{Mambrini:2009ad}, but we expect
similar results to hold.}
$U(1)_A$.
 The electroweak symmetry breaking then generates  $Z'ZZ$, $Z'Z\gamma$ and
 $Z'Z'Z$  vertices.

\noindent
From Eq.\ref{Eq:chernsimon} one can deduce the vertices and Feynman rules after the electroweak breaking:

\bea
&&
\Gamma^{\mu \nu \sigma}_{Z'ZZ} (p_3;p_1,p_2)=  2 \alpha_1 s_W ~ \epsilon^{\mu \nu \rho \sigma}(p_1-p_2)_\rho \ ,
\nonumber
\\
&&
\Gamma^{\mu \nu \sigma}_{Z'Z \gamma} (p_3;p_1,p_2)= 2 \alpha_1 c_W ~ \epsilon^{\mu \nu \rho \sigma}(p_2)_\rho \ ,
\nonumber
\\
&&
\Gamma^{\mu \nu \sigma}_{Z'ZZ'} (p_3;p_1,p_2)=  2 \alpha_2 s_W ~\epsilon^{\mu \nu \rho \sigma}(p_2-p_3)_\rho \ ,
\label{Eq:vertex}
\eea
with the obvious notation $s_W =  \sin \theta_W$ and  $c_W =  \cos \theta_W$.

\begin{figure}[!h]
    \begin{center}
   \includegraphics[width=3.4in]{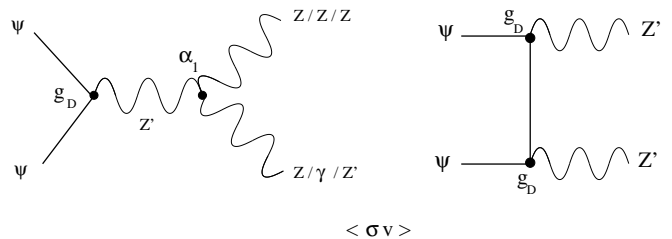}
          \caption{{\footnotesize
Feynmam diagrams contributing to the dark matter annihilation and $\gamma-$ray line observable by FERMI
telescope}}
\label{Fig:feynmanbis}
\end{center}
\end{figure}

\noindent
If $\psi$ is the lighter of the fermions charged under U'(1) (but not under the SM gauge group), coupling to $Z'$  via the vertex
\beq
\Gamma^{Z' \bar \psi  \psi }_{\mu}(p_3;p_1,p_2)= i \frac{g_D}{4} \gamma_{\mu} [(q_L + q_R) +  (q_L - q_R) \gamma^5] \ ,
\eeq

\noindent
where $q_L=q_R=1$ in what follows, it can be considered to be a good dark matter candidate.
The diagrams giving the annihilation rate contributing to the relic abundance are shown in Fig.\ref{Fig:feynmanbis}.
Depending on the kinematics and values of the couplings, each of the two diagrams can dominate.
Nowadays, the dark matter candidate being mainly at rest, the process $\bar \psi \psi \rightarrow Z' \rightarrow Z \gamma$
can generate a monochromatic $\gamma-$ray line observable by the FERMI telescope.
Depending on the mass of $\psi$, we describe in detail each possibility in the following section.

\section*{The monochromatic  $\gamma-$ray line}

Recently, it has been argued that the FERMI telescope did observe a monochromatic $\gamma$-ray line
from the galactic center around a region $E_\gamma \simeq$ 130 GeV, with an annihilation
cross section $\langle \sigma v \rangle \simeq 2 \times 10^{-27} \mrm{cm^3 s^{-1}}$ \cite{Weniger:2012tx},
which would be a tantalizing smoking gun signal for new physics.
Without discussing or anticipating the official analysis of the FERMI collaboration, we will
try to check if such a clear signal can be produced by the CS terms generated in Eq.\ref{Eq:chernsimon}.
In what follows, we need to distinguish three different cases: $2 m_{\psi} < M_{Z'} + M_Z$, $m_{\psi} < M_{Z'} < 2 m_\psi-M_Z$
and $M_{Z'} < m_\psi$, for suitable values of the dark matter mass and couplings to fit the supposed $\gamma$-ray line.
We will show how the first case is strongly disfavored by the astrophysical data and how the latter ones are
 compatible with very natural values for the parameters of the model.
Even if all the discussion in what follows is qualitative, the numerical analysis has been done using
a version of Micromegas \cite{Micromegas}, adapted to include the new features of the model.

\subsection{$M_{Z'} >2 m_{\psi} -M_Z$}

In this case, the only annihilation process kinematically allowed is the $s-$channel exchange of a $Z'$ (see the left Fig.\ref{Fig:feynmanbis}).
However, the two only final states being $ZZ$ and $Z\gamma$, from the values of the couplings in Eq.\ref{Eq:vertex} one can easily
deduce $\frac{\langle \sigma v \rangle_{ZZ}}{\langle \sigma v \rangle_{Z \gamma}}\simeq 0.3$ which means that the $Z \gamma$
final state is always the dominant one. If one wants to fulfill WMAP constraints for a thermal relic
($\langle \sigma v \rangle \simeq 3 \times 10^{-26} \mrm{cm^3 s^{-1}}$), one should impose
$\langle \sigma v \rangle_{Z \gamma} \simeq 2 \times 10^{-26} \mrm{cm^3 s^{-1}}$.
Such a huge cross section would have produced a visible monochromatic line, which is already apriori excluded by the FERMI collaboration.
Therefore, the mass range $2 m_\psi < M_{Z'}+M_Z$ is disfavored.

\subsection{$m_{\psi} < M_{Z'} < 2 m_{\psi}-M_Z$}

\begin{figure}[!h]
    \begin{center}
   \includegraphics[width=3.in]{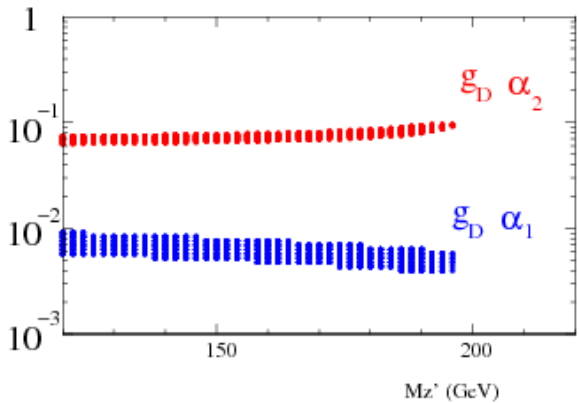}
\vspace*{-2mm}
          \caption{{\footnotesize
Combinations of Chern-Simon coefficients $g_D \times \alpha_1$ and $g_D \times \alpha_2$
respecting WMAP constraint and producing a monochromatic $\gamma-$ray line
around 130 GeV with $\langle \sigma v \rangle_{Z\gamma} \simeq 2 \times 10^{-27}$ \cite{Weniger:2012tx}
for $ M_{Z'} \lesssim 2 m_{\psi} - M_{Z}$ (case B, see the text for details).
}}
\label{Fig:scanbis}
\end{center}
\end{figure}

In this case, the opening of a new channel $\bar \psi \psi \rightarrow Z' \rightarrow Z' Z$ allows
the possibility to obtain a relic density obeying WMAP constraint $and$ a monochromatic $\gamma-$ray
 flux observable by FERMI at the same time. In fact, the main difference with the case discussed in the previous subsection consists in the fact
that now the contribution to the $\bar \psi \psi \rightarrow Z'Z$ process is proportional
 to the second Chern Simons coefficient of the Eq.\ref{Eq:chernsimon}. Therefore, it is possible to
 decouple the two different processes, the annihilation cross section to fulfill WMAP, proportional to $\alpha_2$,
 and the annihilation cross section giving a monochromatic signal, proportional to $\alpha_1$.

 The result is shown in Fig.\ref{Fig:scanbis}, where we plotted the regions of the parameter space
 still allowed by WMAP and respecting
 $0.4 \times 10^{-27} \mrm{cm^3s^{-1}} < \langle \sigma v \rangle_{Z\gamma}< 1.09 \times 10^{-27} \mrm{cm^3s^{-1}}$
 for a dark matter mass $m_{\psi}=144.5$ GeV.
 One can see that the dependence on the parameters can be expressed in terms of the products $g_D \times \alpha_i$. At the same time,
 as the WMAP constraints require a relatively larger cross section than the one for the monochromatic line,
 we require a ratio $\simeq$ 10 between $\alpha_2$ and $\alpha_1$. Therefore, if one consider
 reasonable values for standard coupling of the Chern Simons terms, $\alpha_2 \simeq 10^{-2}$, from Fig.\ref{Fig:scanbis} one obtains $g_D \simeq 1$ and even stronger (non-perturbative) values for smaller values of  $\alpha_2$. This case is therefore compatible with the data,
though a small hierarchy between $\alpha_1$ and $\alpha_2 $ has to be assumed.

\subsection{$M_{Z'}<m_{\psi}$}

In this region the main contribution to the dark matter annihilation process comes from the t-channel $\psi$ exchange depicted in the right panel of Fig.\ref{Fig:feynmanbis}. In this case, the relic density condition (independent on $\alpha_1$ or $\alpha_2$) is essentially decoupled from the $s-$channel
diagram which produces the monochromatic line (Fig.\ref{Fig:feynmanbis} left). We recall again that it was because the $same$
diagram was responsible for the relic abundance and the monochromatic line that for example the region $M_{Z'} >2 m_{\psi} -M_Z$
has been excluded in the discussion above. Moreover, interestingly in this case, differently from the scenario in which $m_{\psi} < M_{Z'} < 2 m_{\psi}-M_Z$, our analysis becomes independent on the parameter $\alpha_2$ as soon as $\alpha_2 \lesssim g_D$.

\begin{figure}[!h]
    \begin{center}
   \includegraphics[width=3.in]{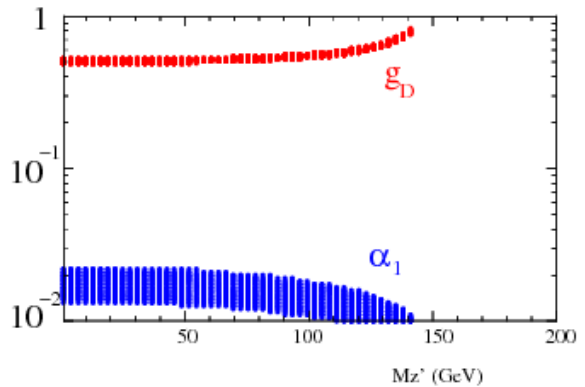}
\vspace*{-2mm}
          \caption{{\footnotesize
Dark coupling $g_D$ and Chern-Simon coefficient $\alpha_1$
respecting WMAP constraint and producing a monochromatic $\gamma-$ray line
around 130 GeV with $\langle \sigma v \rangle_{Z\gamma} \simeq 2 \times 10^{-27}$ \cite{Weniger:2012tx}
for $M_{Z'} \lesssim m_{\psi}$ (case C, see the text for details).
}}
\label{Fig:scan}
\end{center}
\end{figure}

\begin{figure}[!h]
    \begin{center}
   \includegraphics[width=3.in]{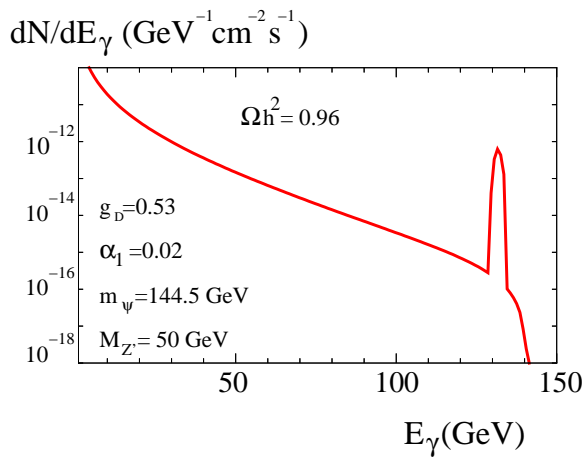}
          \caption{{\footnotesize
Example of spectrum observed from a cone within 0.1 radian from the Galactic Center,
 respecting WMAP and producing a monochromatic $\gamma$-ray line
 around 130 GeV with $\langle \sigma v \rangle_{Z\gamma}\simeq 2 \times 10^{-27}$.
}}
\label{Fig:spectrum}
\end{center}
\end{figure}

\noindent
In more details, for a given monochromatic line (and so a given $m_\psi$), the WMAP condition fixes the coupling $g_D$:
the cross-section for $\bar \psi \psi \rightarrow Z' Z'$ depends only weakly -- through a phase space coefficient -- on $M_{Z'}$.
Then for a given value of the monochromatic annihilation cross section $\langle\sigma v \rangle_{Z \gamma} $, one
can deduce the value of $\alpha_1$ fitting the FERMI data. We made a scan on ($M_{Z'}$, $g_D$, $\alpha_1$)
and applied the $5 \sigma$ constraint from WMAP and the annihilation cross section proposed by \cite{Weniger:2012tx}
in the case of an Einasto profile. The result is presented in Fig.\ref{Fig:scan}.
In Fig. \ref{Fig:spectrum} we present an example of the gamma-ray spectrum obtained from  the diagrams 
shown in Fig. \ref{Fig:feynmanbis}, for a  point in the parameter space
respecting WMAP\footnote{Micromegas was used for the calculation.}.

\noindent
There are three striking features in Fig.  \ref{Fig:scan}: first, there is a weak dependence on $M_{Z'}$; secondly, the
value of $g_D$ takes natural value for a $U(1)$ coupling (we remind that $g_{EW}=0.65$);
and finally $\alpha_1$ takes typical one-loop order values $\simeq 10^{-2}$, which are quite consistent with loop
contributions generated by triangle diagrams of Fig.\ref{Fig:feynman}.

\noindent
This range of values is understandable. Indeed, we know that an annihilation cross section
$\langle \sigma v \rangle \simeq 3 \times 10^{-26}\mrm{cm^3s^{-1}}$ leads to values of couplings of the order of the electroweak one for a WIMP mass of 100 GeV. As only the dark coupling $g_D$ appears in the annihilation channel (Fig.\ref{Fig:feynmanbis} right),
one expect $g_D \simeq$ 0.6 for a WIMP mass of $\sim$ 100 GeV independently on $Z'$ mass (except around the threshold)
as one can see in Fig.\ref{Fig:scan}. Now, if one imposes that
$\langle \sigma v \rangle_{Z\gamma}\simeq 10^{-27}\mrm{cm^3s^{-1}}$, one can check semi-analitically that
$\alpha_1 \simeq g_D/30$ which is effectively what we also observe in Fig.\ref{Fig:scan}.

\noindent
It could be interesting to notice that in any of the cases discussed above, the direct detection rate is largely suppressed. Indeed,
this rate can become important for a kinetic mixing around $\delta \simeq 10^{-3}$ \cite{Mambrini:2010yp}
and could even explain DAMA/CoGENT excess with $\delta \simeq 10^{-2}$. However, in the model we are considering, the ``portal"
between the dark matter sector and the visible one does not go through this kinetic mixing, but through the tri-vectorial
couplings generated in Eq.\ref{Eq:chernsimon}. In this case, very low values of $\delta$ are still allowed,
rendering the direct detection (through t-channel Z' exchange) very difficult to observe.

\noindent
One can also remark that for $M_{Z'} \lesssim M_{Z}$, the $Z\gamma$ final state channel is kinematically closed.
The main decay channel for the $Z'$ is thus through the kinetic mixing with the $Z$.
It is important to check that in this case the kinetic mixing should not be too small to disturb the Big Bang Nucleosynthesis
(BBN)
problem. A straightforward computation of the $Z'$ lifetime leads to

\bea
&&
\Gamma_{Z'\rightarrow q \overline q} \simeq \frac{e^2 \delta^2 \cos^2 \theta_W M_{Z'}}{108 \pi}
\\
&&
\Rightarrow \tau_{Z'} \simeq 10^{-22} \frac{8 \pi}{e^2 \delta^2 \cos^2 \theta_W}\left( \frac{M_{Z'}}{1 GeV} \right) \
(\mrm{seconds})~~ \ .
\nonumber
\eea

\noindent
One may deduce that for $\delta \gtrsim 10^{-11}$, the $Z'$ decays before one minute and will not affect the BBN processes.
Thus, the region $10^{-11} \lesssim \delta \lesssim 10^{-2}$ allows for a safely $Z'$ decay from the BBN and electro-weak precision tests
point of view. Moreover, notice that the natural one-loop contribution to the mixing, due to the diagrams induced by CS interaction terms,
stays well inside that windows, for interesting (one-loop) values of the couplings $\alpha_i $ and reasonable assumptions\footnote{Stronger constraints on $\Lambda$ are discussed in the conclusion.} for the scale $\Lambda$ where the theory is completed (for example, the mass $M$ of the heavy fermions).

\noindent
One should also notice that several constraints from the continuum photons should be taken into account,
especially the ones coming from the dwarf galaxies measured by FERMI telescope\footnote{We are grateful to the referee for
having pointing us this issue.} \cite{FERMIDWARF}. Indeed, one have checked that the photons fluxes generated by the subsequent decays 
of the $ZZ'$ final state (case B) or $Z'Z'$ (the $ZZ$ final state giving weak fluxes, reduced by a factor $\simeq \sin^2 \theta_W$ compared
to $Z \gamma$) does not exceed the constraints obtained by FERMI collaboration \cite{FERMIDWARF}.



\section*{Conclusion}

We have discussed an extension of the Standard Model with an extra $U'(1)$ abelian group, where a
three gauge boson couplings $Z'Z\gamma$ is generated from Chern-Simons terms. We studied the different scenarios allowed by this model, 
for different values of the mass parameters, under the hypothesis that the dark matter candidate is charged only under the extra abelian group.
 Depending on the ratio between the mass of the dark matter and the mass of the mediator $Z'$, WMAP data constrains, more or less severely, 
 the gauge coupling of the group $g_D$, but always allowing it to have a very natural value comparable with the usual electroweak ones, 
 independently of the absolute value of the $Z'$ mass. At the same time, for $M_{Z'} < 2 m_{\psi}-M_Z$ and $M_{Z'} < m_{\psi}$
the model can provide a monochromatic $\gamma-$ray line which can fit a 130 GeV signal at FERMI telescope for a dark matter candidate
 mass $m_{\psi}=144.5$ GeV, again for rather natural values of the Chern-Simons  couplings.

 \noindent
 Obviously, the model presented is intended to be an effective theory, where the unique effects of the beyond the standard model physics are 
 encoded in the trilinear vector bosons couplings and the presence of a fermionic dark matter candidate.
  Nonetheless, already at this level it is possible to have a good estimate of the constraints that the new physics should satisfy in order to fit with 
  this dark matter scenario.
  For example, the CS term in (\ref{Eq:chernsimon}) contributes, at the loop level, to the mass of the $Z$ gauge boson, which is experimentally
known with an accuracy of $2-3$ MeV. An order of magnitude estimate gives $\delta M_Z^2 \sim (\alpha_i^2/16 \pi^2) \Lambda^2$, 
where again $\Lambda$ is an UV cutoff naturally of the order of the fermions generating the CS term. For couplings of 
order $\alpha_i \sim 10^{-2}$, we find $\Lambda \leq 500$ GeV, which is marginally consistent with limits on vector-like fermions in the 
Standard Model. It would be interesting to compare the constraints we obtained on  $\alpha_1$ and $g_D$ from our combined WMAP/FERMI 
analysis with the constraints one could find with the LEP searches through the process
 $e^+e^- \rightarrow Z \rightarrow \gamma Z' \rightarrow \gamma \gamma Z$
  which is a $2 \gamma$ plus $Z$ final state signature. However, such analysis is  beyond the scope of the present paper.

\vskip1cm

\noindent {\bf Acknowledgements. }  The authors would like  to thank E. Bragina, A. Morselli, G. Zaharijas, G. Gomez
and the Magic Monday Journal Club for discussions. This  work was supported by the European ERC Advanced Grant 226371
MassTeV, the French ANR TAPDMS ANR-09-JCJC-0146  and the Spanish MICINNÕs Juan de la Cierva  and
Consolider-Ingenio 2010 Programme  under grant  Multi-Dark CSD2009-00064 and CPAN CSD2007- 00042.
We also thank the support of the MICINN under grant FPA2010-17747, the Community of Madrid under grant
HEPHACOS S2009/ESP-1473, and the European Union under the Marie Curie-ITN programme
PITN-GA-2009-237920.
S.P. acknowledges partial  support of the TUM-IAS funded by the German
Excellence Initiative,
by the contract PITN-GA-2009-237920 UNILHC and by
the National Science Centre in Poland under research grant
DEC-2011/01/M/ST2/02466

%
%
%
%
%
%
%
%
%
%
%
%
%
%
%
%
%
%
%
%
%
%
%
%
%
%
%
%
%
%


\end{document}